\documentclass[12pt]{article}

\usepackage{graphicx}

\begin{document}
%\titlerunning{Title running}
\begin{center}
{\Large\bf \boldmath ON SOME FEATURES OF COLOR CONFINEMENT} %<== title (bold face, capitalize)

\vspace*{6mm}
{Adriano Di Giacomo} \\      %<== authors
{\small \it  Pisa University and INFN Sezione di Pisa}      %<== institutions
 \end{center}

\vspace*{6mm}

% abstract
\begin{abstract}
 It is argued that a dual symmetry is needed to naturally explain experimental limits on color confinement. Since color is an exact symmetry the only possibility is that this symmetry be a dual symmetry, related to non trivial spatial homotopy. The sphere at infinity of 3-dimensional space being 2-dimensional, the relevant homotopy is $\Pi_2$ , the corresponding configurations monopoles, and the mechanism  dual superconductivity. The consistency of the order-disorder nature of the deconfining transition is compared with lattice data . It is also shown that the only dual quantum number is magnetic charge and the key quantity is 't Hooft tensor, independent of the  gauge group. The general form of the
 't Hooft tensor is computed. 
 \end{abstract}

\vspace*{6mm}

\section{Introduction}

Experimental upper limits to the observation of free quarks in Nature are very stringent \cite{PDG}.
 Typically for the abundance of quarks in ordinary matter $n_q$ as compared to that of protons $n_p$
 the limit is ${n_q/n_p}\le 10^{-27}$ to be compared to the expectation in the Standard Cosmological Model in absence of confinement  ${n_q/n_p}\approx 10^{-12}$\cite{Okun}. 
 
 The natural explanation is that $n_q$
is exactly zero due to some symmetry . In this case the deconfining transition is an order disorder transition and can not be a crossover. A crossover indeed means continuity and the theory should explain 
a factor of $10^{-15}$ for $n_q$ for a continuous transition between  the two phases.  

This is similar to what happens in ordinary superconductivity , where a very small upper limit is observed for the resistivity  $\rho_{SC}$  in the superconducting phase with respect to the normal one .
There the transition is from a Higgs broken phase ( superconductor ), in which $\rho_{SC}$ is strictly zero
for symmetry reasons, to a Coulomb phase (normal) in which electric charge is superselected.

 If this  argument is correct two main questions raise naturally, namely :
  
  1) What symmetry is responsible for confinement ?
  
  2) Is an order disorder transition compatible with observation?
  
  No clear observation exists yet of deconfinement in heavy ion experiments. There is a clear evidence of it, however, in simulations of the theory on a Lattice from first principles. There the  deconfining transition can be observed and  its order and universality class can be determined, at least in principle.
  
  \section{Symmetry}
  
  Color is believed to be an exact (Wigner) symmetry . Perturbative  $QCD$ is  based on BRST symmetry, which is nothing but the statement that vacuum is a color singlet.  Therefore color can not distinguish the confining phase from the deconfined one.  What can then be an extra symmetry, besides color, which can do that?
  
  In quenched $SU(N)$ gauge theory (no dynamical quarks) there is a cheap answer : the center of the group , $Z_N$.  The Lagrangean of pure gauge theory is indeed blind to $Z_N$, since gluons belong to the adjoint representation.  Usually, however , the theory is formulated on the lattice in terms of parallel transports in the fundamental representation, to allow the introduction of static quarks as external sources. $Z_N$ is then a symmetry of the particular regularization. Static quarks are described by a parallel transport along the time axis, the Polyakov line $L(\vec x)$.
  \begin{equation}
      L(\vec x) =  P \exp [ i  \int  _0^{1\over T} A_0(\vec x, t) dt]
  \end{equation}
  The lattice is supposed to be extended in the time direction from zero to the inverse of the temperature $T$.
   The static potential acting between a static   $q \bar q$ pair,  $V(\vec x) $ , is related to Polyakov loop correlators as
   \begin{equation}
   V(\vec x) = -T  ln( \langle L^{\dagger}(\vec x) L(\vec 0) \rangle)
   \end{equation}
   Since, by general arguments, the cluster property holds 
   \begin{equation}
   \langle L^{\dagger}(\vec x) L(\vec 0) \rangle \approx  |\langle L \rangle|^2+ C \exp (-{\sigma x \over T})
   \end{equation}
   when   $\langle L \rangle =0$     $V(\vec x)\approx_{x\to\infty}  \sigma x$  ( confinement)
   
   when   $\langle L \rangle \neq 0$      $V(\vec x)\approx_{x\to\infty}$ constant (deconfinement).
   
    $\langle L \rangle $  is the order parameter and the symmetry is $Z_N$.
    
  However in Nature  dynamical quarks do  exist and their coupling explicitly breaks  $Z_N$, which then cannot be the symmetry responsible for  confinement.
  
  In front of this difficulty there exist in the community two different attitudes:
  
  a)  A narrow minded, conservative attitude : the only extra symmetry is a flavor symmetry, namely the chiral symmetry at zero quark mass.
  
  b) A more advanced attitude looking for a dual symmetry related to topologically non trivial spatial boundary conditions. \cite{KW} \cite{KC}\cite{'tH75}\cite{m75}\cite{SW}\cite{GW}.
  
  Some comments on the attitude  a).  If the only relevant degrees of freedom at the deconfining  transition at $m_q=0$ $N_f =2$ are the chiral ones then a renormalization group argument leads to the conclusion that  either the transition is second order and belongs to the universality class of $O(4)$ in 3-d , and in that case the transition at  $m_q\neq 0$ around the chiral point is a crossover. Or it is first order and then also at small non zero masses it is first order\cite{PW}. The first possibility is very popular in the literature \cite{owe}, but nobody has found consistency of  Lattice data with the $O(4)$ critical indexes.
  More recent data\cite{ddp}\cite{cddp}, instead, show consistency with a weak first order.
  Moreover, if chiral degrees of freedom were the only relevant ones, one should expect that also in the analogous system with 2 flavors of quarks in the adjoint representation, they should dominate. Instead that system shows two different transitions \cite{K}\cite{CDDLP}  : a strong first order deconfining transition which is detected, e.g. by the Polyakov line ( $Z_3$ is a symmetry  for adjoint quarks) and a very weak chiral transition which is consistent with a cross-over. This demonstrates  that there exist other relevant degrees of freedom than light  chiral scalars at the deconfining transition.
  \section{Duality}
  The key word for the approach b) is {\bf duality}.
   The prototype example is the 2-d Ising model. It can be viewed as the discretization of a (1+1)-dimensional field theory, the field being the variable $\sigma = \pm1$ defined on each site of a two dimensional square lattice. The partition function is that of a paramagnetic nearest-neighbours $i, j$ interaction  $ Z=  \Sigma _{i,j} \exp ( - { \beta} \sigma_i \sigma_j) $. The system has a second order phase transition from an ordered phase $\langle \sigma \rangle  \neq 0$ to a disordered phase $\langle \sigma \rangle  = 0$ at a critical value $\beta_c$. The system admits spatial 1-dimensional configurations with non trivial topology, the kinks (anti-kinks), with $ \sigma = -1(+1)$ for $x \leq x_0$ and 
  $\sigma = +1(-1)$ for $x  > x_0$ .  The operator $\mu$ which creates a kink at  a given time reverses the sign of $\sigma$ for $x \le x_0$ and time $t$. If it acts twice the result is the identity, so that $\mu^2 =1$ or  $\mu = \pm 1$ . It is a theorem that\cite{KC}

  \begin{equation}
  Z [ \sigma, \beta]  = Z [ \mu, \beta^*]
  \end{equation}
  \begin{equation}
    sinh(2 \beta) = {1\over sinh(2 \beta^*)}  
    \end{equation}
    or                                $\ beta \approx {1/over \beta^*}$.\\
 It follows that below  the critical temperature   $\langle \sigma \rangle  \neq 0$  and  $\langle \mu \rangle  = 0$ ,
 above it   $\langle \sigma \rangle  =  0$ and  $\langle \mu \rangle  \neq 0$.
 A topological current $j_{\mu}$ can be defined which is identically conserved $\Delta_{\mu} j_{\mu} = 0$
 \begin{equation}
j_{\mu}  = \epsilon_{\mu,\nu} \Delta_{\nu} \sigma
\end{equation}
The corresponding conserved charge is  $Q = \int  dx j_0(x,t) = \sigma(+\infty) -\sigma(-\infty)$ Is equal to the number of kinks minus the number of anti-kinks.
 In summary the system admits two equivalent descriptions [eq(4)]: either in terms of the local fields $\sigma$,
 and in this description the topological excitations are non local (direct description) , or in terms of the dual variables $\mu$
 as local variables, and then the fields $\sigma$ are non local (dual description). The first one is convenient at low temperature, the second one at high temperature. Duality maps the weak coupling regime of the direct description into the strong coupling of the dual, and viceversa.
 
  In (3+1)dimensional field theories dual configurations have non trivial $\Pi_2$ corresponding to a non trivial mapping of the 2-dimensional sphere at infinity on the fields , and are monopoles.
  In (2+1) dimensions the dual configurations  have non trivial $\Pi_1$ and are vortices.
  
  \section{Monopoles}
  Monopole configurations in non abelian gauge theories were first studied in a Higgs model with gauge group $SU(2)$ and Higgs field in the adjoint representation \cite{'tH}\cite{Pol} . Everything is in the adjoint representation so that the theory is blind to $Z_2$ and is in fact an $SO(3)$ gauge theory.
  Monopoles of ref's\cite{'tH}\cite{Pol} are static classical solutions of the equations of motion with finite energy (Solitons). In the "hedgehog" gauge the Higgs field has the form
  \begin{equation}
  \phi^a (\vec r) = f(r) {r^a\over r}  
  \end{equation}
  The orientation of the field in color space coincides with that of the position vector in physical space.
  $f(r) \to 1 $ as $r \to \infty $ and therefore the solution is a non trivial mapping of the two-dimensional sphere at spatial infinity onto the group $SO(3)/U(1)$ , $U(1)$ being the invariance group of $\vec \phi$.
  From the general formula
  \begin{equation}
  \Pi_2 (G/H) = ker[\Pi_1(H)  \to \Pi_1(G)]
  \end{equation}
  valid for any breaking of a group $G$ to a subgroup $H$ we get  \cite{DLP}  $\Pi_2[SU(2)/U(1)] = \Pi_1[U(1)] = Z$  and  $\Pi_2 [SO(3)/U(1)] = Z/Z_2$ .  In the model of ref.\cite{'tH}\cite{Pol} configurations are labeled by an even integer.  This integer is nothing but the magnetic charge in units of Dirac units ${1\over 2g}$ with $g$ the gauge coupling constant which plays the role of electric charge.
  
   Since a monopole is always an abelian configuration \cite{Coleman} the magnetic charge has to be 
   coupled to the residual $U(1)$ gauge group, i.e. to the abelian field  strength
   \begin{equation} 
   F_{\mu \nu} = \partial_{\mu} A^3_{\nu} - \partial_{\nu} A^3_{\mu}
   \end{equation}
 where $ A^3_{\mu}$ is the projection of the gauge field  along the Higgs field ${\vec  \Phi}$ in the unitary gauge in which it is directed along the third axis.
 
The tensor $ F_{\mu \nu} $ can be given a gauge invariant form\cite{'tH}, which is known as 't Hooft tensor.  Denoting by $\vec \phi =\vec \Phi/ |\vec \Phi/|$ the direction of the Higgs field in color space 
   one can show that \cite{'tH}
   \begin{equation}
   F_{\mu \nu} = \vec \phi.\vec G_{\mu \nu} - {1\over g}  \vec \phi.( \vec {D_{\mu} \phi} \wedge \vec {D_{\nu} \phi})
   \end{equation}
   We define the current
   \begin{equation}
   j_{\nu} =  \partial_{\mu}  F^*_{\mu \nu}
   \end{equation}
   with the usual notation for the dual tensor $F^*_{\mu \nu}= {1\over 2}\epsilon_{\mu \nu \rho \sigma}F_{\rho \sigma}$ . Normally, with trivial boundary conditions $\partial_{\mu}  F^*_{\mu \nu} =0$ an equality known as Bianchi identity. For  $ j_{\nu} \neq 0$  one always has 
   \begin{equation}
   \partial_{\nu} j_{\nu} =0
   \end{equation}
   This is a topological symmetry, not related to the action via Noether's theorem and is our dual symmetry.
   The corresponding conserved charge  is the magnetic charge $Q= \int d^3x j_0(\vec x, t)$.
   For the monopole configuration   one has
   \begin{eqnarray}
   E_i &=& F_{o i} =0\\
   \vec H &=& {1\over g} {{\vec r}\over r^3} + Dirac -string
   \end{eqnarray}
   In a compact formulation like lattice the Dirac string is not visible so that    $j_0 = \vec \nabla \vec H ={ (4 \pi) \over g} \delta^3(\vec r)$ , a violation of Bianchi identity. The monopole is a configuration of charge 2
   in agreement with the geometric argument above. The charge $Q$ labels the dual degrees of freedom.\\The presence of the Higgs field is only necessary if one wants the monopole as a soliton and a real breaking of the symmetry. In fact the role of $\vec \Phi$ can be played by any operator $\Psi$ in the adjoint representation : monopoles will be located at the zeroes of $\Psi$ , and their number and location will depend on the choice of  $\Psi$ , but a conserved current will be always defined as in eq(11).
   However one can think of a theory defined everywhere
   in space time, except for a discrete but arbitrary number of line like singularities which describe the
   dual degrees of freedom [ Witten's geometric Langland's program\cite{GW}].  Creating a new monopole
   by an operator $\mu$ means adding a new singularity and this will be true whatever the choice 
   of $\Phi$ .  The vacuum expectation value $\langle \mu \rangle$ will be zero if the magnetic charge is
   super-selected and the vacuum has a definite magnetic charge. If, instead, $\langle \mu \rangle\neq 0$ magnetic gauge symmetry is broken a la Higgs and the vacuum is a dual superconductor.
    $\langle \mu \rangle$  can be used as an order parameter for confinement.\\
    The above construction can be extended to any gauge group coupled to any matter fields: the basic ingredients are indeed gauge symmetry and the fact that physical space is 3-dimensional.
    For a generic gauge group there are $r$ independent magnetic currents, with $r$ the rank of the group\cite{DLP}. The corresponding effective Higgs fields are the fundamental weights of the group.
    The 't Hooft tensor corresponding to each of them can be explicitly computed\cite{DLP} and has a more complicated form than that of Eq(10). The residual symmetry can be immediately read from the Dynkin diagram of the Lie algebra, and is the Levy subgroup obtained by eliminating the little circle 
    of  the simple root corresponding to the given fundamental weight \cite{DLP}.
    \section{The order parameter for confinement}
    An order parameter for monopole condensation has been developed mainly in Pisa in recent years
    to detect dual superconductivity of $QCD$ vacuum as explained above\cite{d} \cite{dd}\cite{cc}\cite{ddd}. The idea is to translate the component of the gauge field along the residual $U(1)$ direction
    by a classical monopole field by use of the conjugate momentum.  In formulae
    \begin{equation}
    \mu^a(\vec x, t)  = \exp [i \int d^3y {m\over g}\vec E^a(\vec y,t) \vec b_{ \perp} (\vec x - \vec y)] 
    \end{equation}
    ${m\over g}$ is the magnetic charge of the monopole,  $\vec b_{\perp} (\vec z ) = {{ \vec n \wedge \vec z}\over {z(z-\vec n.\vec z)}}$ is the vector potential of the field generated by it in the transverse gauge  $\vec \nabla \vec b =0$  $\vec \nabla \wedge \vec b_{\perp}(\vec z) = {{\vec z}\over z^3} $+
    Dirac - String along the direction $\vec n$.\\ $\vec E^a = Tr [ \Phi^a \vec E]$ with $\Phi^a$ the $a-th$ fundamental weight  is the component of the Chromoelectric field along the residual $U(1)$ symmetry
 $T^a_3$   coupled to the magnetic charge. In the convolution with $\vec b_{\perp}$ only the transverse part contributes, which is the conjugate momentum to the transverse vector potential $\vec A^{a3}_{\perp}$
    so that
    \begin{equation}
    \mu^a(\vec x, t) | \vec A^{a3}_{\perp}(\vec z, t)\rangle =  | \vec A^{a3}_{\perp}(\vec z, t) +{m\over g}\vec b_{ \perp} (\vec x - \vec z) \rangle
    \end{equation}
    The operator $\mu^a$ simply adds a monopole to the residual abelian projected field.
    It is easy to show that  $\mu^a$ depends on $\beta \equiv {{2N_c}\over g^2}$ as 
    $\mu^a = e^{- \beta \Delta S^a}$ so that
    \begin{equation}
    \langle \mu^a \rangle = { {\int [d\phi] e ^{- \beta (S + \Delta S^a )}}\over  {\int [d\phi] e ^{- \beta S  }}}= {Z(S+\Delta S^a)\over {Z(S)}}
    \end{equation}
    the ratio of two partition functions, which is $1 $ at $\beta =0$.
    
     If we define\cite{dd} \cite{cc}  $\rho^a \equiv {{\partial \ln(\langle \mu^a\rangle)}\over \partial \beta}$,
     we get then
     \begin{equation}
     \langle \mu^a \rangle = \exp( \int d\beta ' \rho^a(\beta ') d\beta')
     \end{equation}
     If a deconfining transition exists at $T=T_c$ , in the thermodynamical limit $V \to \infty$\cite{dd}\cite{cc}
     
     1) $\rho^a \to \bar \rho^a$ a  finite limit     for $T \le T_c$ so that $\langle \mu^a \rangle \neq 0$ (confinement)
     
     2) $\rho^a \propto -V^{1\over 3} \to - \infty$  for $T > T_c$ so that  $\langle \mu^a \rangle = 0$ (deconfinement)
     
     3) At $T \approx T_c$ $\langle \mu^a \rangle$ drops to zero and therefore  $\rho^a$ has a negative peak,
      which signal the transition.  Moreover the  finite size scaling dependence on the spatial size of 
      the system $L_s$ ($L_s^3 =V$) is 
      \begin{equation}
      \rho^a = L_s^{1\over \nu} \phi^a( \tau L_s^{1\over \nu})
      \end{equation}
      where $\tau = 1 - {T\over T_c} $ is the reduced temperature, and $\nu$ the usual critical index
      of the correlation length at the transition.
       Not only $\rho^a$ (or  $\mu^a$) detects the transition , but it also provides information on its order and  universality class .
       
       The behavior described above has been checked in a number of systems, in particular to
       study the phase diagram of $N_f=2$ $QCD$ at small quark masses\cite{fqcd}.
       In the physical case where the quarks are in the fundamental representation the deconfining transition coincides with the chiral transition: the negative peak of $\rho^a$ seats just at the temperatore where 
       the chiral order parameter $\langle \bar \psi \psi\rangle $ drops to zero. The scaling Eq(19) is compatible with a weak first order transition. In the case of the quarks in the adjoint representation of the color group the deconfining transition takes place at lower temperature than the chiral one, it is detected by $\rho^a$ ($\mu^a$) and is consistent with first order\cite{CDDLP}. The chiral restoration is instead consistent with a very weak  crossover.
     \section{Order-disorder and Lattice}
     The deconfining transition is popularly believed to be a crossover in a wide region of the $QCD$ phase diagram[ See e.g. ref.\cite{owe} for a review].  This only means that no evident jump of any physical
     quantity has been detected up to presently available volumes. In principle it is not possible to state on the basis of data [ numerical or experimental] that a transition is a crossover and not a weak first order: the only correct statement can be " this transition is consistent with a crossover up to the presently available volumes". 
     There are cases , however, in which theoretical arguments allow to state that there is a crossover.
     For example in $N_f=2$ $QCD$ if the chiral transition at $m_q=0$ is second order then by general arguments at small masses in the neighborhood of $m_q=0$ the transition will be a crossover.
    [ Notice that we are here using a rather improper language by calling transition a crossover.]
     We have already touched this question in Section 2 above. It is therefore very important to
     check if the chiral transition is first order or second order in the universality class of $O(4)$.
     In the first case the transition is first order also at $m_q\neq 0$ , a scenario compatible with
     an order-disorder nature of the deconfinement transition. If, instead the chiral transition proves to be second order then it becomes a crossover at $m_q\neq 0$, and order-disorder is ruled out.
      In the second case there is no way to define confinement and deconfinement\cite{ddp}\cite{cddp} .
     This analysis can be done by use of finite size scaling techniques: the behavior of quantities like  e.g. the specific heat or the susceptibility of the order parameter with increasing volume is governed
     by the critical indexes which are characteristic of the order and universality class of the transition.
     No consistency has been found of the scaling with second order $O(4)$ \cite{owe}.
     In ref.\cite{ddp} new tools of investigation were introduced with respect to the previous literature:
     for example the specific heat, which is independent on any prejudice on the symmetry, was used besides the chiral susceptibility, and a better determination of the reduced temperature including its dependence on the quark mass.  The scaling law for the specific heat reads
     \begin{equation}
     C_V- C_0  = L_s^{\alpha \over \nu} \Phi_C( \tau L_s^{1\over \nu }, m_q L_s^{y_h})
     \end{equation}
     $C_0$ is a subtraction corresponding to a quadratic divergence, which is  ultraviolet and hence 
     independent on the volume and on the quark mass. 
     The critical index $\alpha$  is equal to 1 for a weak first order transition , whilst for second order $O(4)$  $\alpha = -0.2$ , ${ \nu} = {1\over 3}$  for weak first order, $ {\nu} = .748(14)    $     for second order $O(4)$,  $y_h =3$ for weak first order and $y_h= 2.48        $   for second order $O(4)$. In the analysis of Ref\cite{ddp}\cite{cddp}
     first $C_0$ was determined and verified to be independent on $L_s$ and on $m_q$.
     Then a number of simulations were made in which one  the two variables of the scaling function $\Phi$
     of eq(20) was kept fixed in turn assuming either $O(4)$ or weak first order. Keeping the second variable fixed while varying $m_q$ and $L_s$ with the appropriate value of $y_h$, the scaling in the 
     other variable can be tested. In particular at the maximum $C_V- C_0  \propto L_s^{\alpha \over \nu}$
      The scaling is compatible with first order and definitely excludes $O(4)$ : indeed for $O(4)$ $\alpha$
      is negative, but the peak strongly increases wit $L_s$. Keeping the first variable fixed instead, if the transition is first order  one expects \cite{cddp} at large volumes
      \begin{equation}
     C_V- C_0  =  m_q^{-1} \phi^{(1)}(\tau V) + V \phi ^{(2)}(\tau V)
     \end{equation}
     The first term is non singular in the thermodynamical limit, the second term is singular and
     produces a latent heat . In the case of second order instead there is no singularity and only the analog of the first term is present\cite{ddp} namely
     \begin{equation}
     C_V- C_0  =  m_q^{{-\alpha}\over {\nu y_h}} \phi(\tau L_s^{1\over \nu}) 
     \end{equation}
      The importance of the second term increases with the volume and becomes dominant at large enough volumes. For  very weak first order transitions the first term is 
     dominant up to large volumes. The present situation with $N_f=2$ $QCD$ is that the first term is still big \cite{ddp}\cite{cddp}, and scales with the indexes of weak first order, i.e. as the first term of eq(21): the second term is there
     but is not dominant at present volumes.
     More work is needed to give a final clear answer to the question.
     \section{Conclusions}
     The only way to have an operative definition of confinement and deconfinement is to have a symmetry to distinguish them . This also appears to be the natural explanation of the strict upper limits 
     on the observation of free quarks in nature.
      Color is an exact symmetry, and hence the possibility of a symmetry governing  confinement relies on duality.
      This means excitations with non trivial $\Pi_2$, i. e. monopoles. Dual superconductivity is then the candidate mechanism for confinement. This is independent on the gauge group and on the specific matter fields coupled to it.
      An order parameter can be defined for the dual symmetry, which is the expectation value of an operator which carries magnetic charge.
      Lattice data support this scenario in $N_f=2$ $QCD$ and seem to exclude $O(4)$ second order chiral transition, which would imply a crossover at $m_q \neq 0$ which is not compatible with order-disorder transition. More work is needed to definitely clarify the issue.
    
% figure
%Figures should be prepared in the Encapsulated Postscript format and
%attached to the manuscript using \texttt{zip} or \texttt{tar.gz}
%compression files.
%\begin{figure}[h]
%\centerline{\includegraphics[width=7cm]{spmtp08.eps}}
%\caption{Figure caption.}
%\end{figure}
%tables
%Tables may be prepared using either the conventional \texttt{table}
%environment or with the help of any of the standard \LaTeX\ packages.
%{\bf Important!} They should fit the page size!
% references


\begin{thebibliography}{99}\itemsep -1mm

\bibitem{PDG}
Review of Particle Physics ,EPJ{\bf15},
  (2000)
  
  \bibitem{Okun} 
 L. Okun , Leptons and Quarks,North Holland (1982)

\bibitem{KW}
H.A, Kramers, G.H. Wannier
Phys. Rev.{\bf 66},252 (1941)

\bibitem{KC}
L.P. Kadanoff, H. Ceva
Phys.Rev{\bf B3}, 3918 (1971)

\bibitem{'tH75}
 G.'tHooft,  in $High Energy Physics$, EPS International Conference, Palermo 1975, A. Zichichi ed.
 
  \bibitem{m75}
 S. Mandelstam, Phys. Rep.{\bf 23C}, 245 (1976)

\bibitem{SW}
N. Seiberg, E. Witten
Nucl.Phys. {\bf B341}, 484 (1994)

\bibitem{GW}
S.Gukov, E.Witten,
 Gauge theory, ramification, and the geometric Langlands program
  arXiv:hep-th/0612073.

\bibitem{PW}
 R.D. Pisarski, F. Wilczek , Phys. Rev.{\bf D29},
 338 (1984)
 
 \bibitem{owe}
Owe Philipsen , Status of Lattice Studies of the QCD Phase Diagram.
 International Symposium Fundamental Problems in Hot and / or Dense QCD, Kyoto, Japan -2008.
arXiv:0808.0672 [hep-ph] 

\bibitem{ddp}
 M. D'Elia,A. Di Giacomo,C. Pica , Two flavor QCD and confinement.
 Phys.Rev.D72:114510,2005.

\bibitem{cddp}
G.Cossu, M. D'Elia,A. Di Giacomo,C. Pica
Two flavor QCD and confinement II.
 arXiv:0706.4470 [hep-lat] 
 
 \bibitem{K}
 F. Karsch, M. Lutgemeier
 Nucl.Phys.{\bf B550}, 449(1999)
 
 \bibitem{CDDLP}
 G. Cossu , M.D'Elia, A. Di Giacomo , G. Lacagnina , C. Pica 
 Phys.Rev.{\bf D77}:074506,2008.  
 
\bibitem{'tH}
  G.'t Hooft,
  Nucl.Phys. B {\bf 79} (1974) 276.

\bibitem{Pol} 
A.M. Polyakov,
  JETP Lett.  {\bf 20} (1974) 194

 \bibitem{Coleman}
 S. Coleman, \emph{Classical lumps and their quantum descendants}
(1975) published in \emph{Aspects of symmetry (selected Erice
lectures),} Cambridge University Press (1985).

 
 
 \bibitem{DLP}
 A. Di Giacomo, L. Lepori, F. Pucci 
 Homotopy, monopoles and 't Hooft tensor for generic gauuge groups.
 arXiv:0808.4041 [hep-lat] 

 \bibitem{d}
A.Di Giacomo 
  Acta Phys.Polon.B25:215-226,1994. 
 
 \bibitem{dd} 
 A. Di Giacomo, G. Paffuti
Phys.Rev.D56:6816-6823,1997. 

\bibitem{cc}
A. Di Giacomo , B. Lucini, L. Montesi, G. Paffuti .
Phys.Rev.D61:034503,2000.

\bibitem{ddd}
A. Di Giacomo , B. Lucini, L. Montesi, G. Paffuti .
Phys.Rev.D61:034504,2000.

 \bibitem{fqcd}
M. D'Elia, A. Di Giacomo, B. Lucini, G. Paffuti, C. Pica. 
 Phys.Rev.D71:114502,2005.

 
\end{thebibliography}
\end{document}